\documentclass[fleqn,usenatbib]{mnras}

\usepackage{newtxtext,newtxmath}
\usepackage{amsmath,marvosym}
\usepackage{multirow}
\usepackage{xcolor}
\usepackage{ulem}
\usepackage{tablefootnote}
\usepackage[T1]{fontenc}

\DeclareRobustCommand{\VAN}[3]{#2}
\let\VANthebibliography\thebibliography
\def\thebibliography{\DeclareRobustCommand{\VAN}[3]{##3}\VANthebibliography}

\usepackage{graphicx,epsfig}	% Including figure files
\usepackage{amsmath}	% Advanced maths commands
\usepackage{tipa}
\title[]{Activity of comet 7P/Pons-Winnecke during the 2021 apparition}

\author[I. Mariblanca-Escalona et al.]{
Irene Mariblanca-Escalona,\thanks{E-mail: irenem@iaa.csic.es}
Luisa M. Lara, 
Fernando Moreno,
Pedro J. Gutiérrez
\newauthor and Marçal Evangelista-Santana 
\\
Instituto de Astrofísica de Andalucía, CSIC, Glorieta de la Astronomía s/n, E-18008 Granada, Spain}

\date{Accepted 2025 February 26. Received 2025 February 26; in original form 2024 December 17}

\pubyear{2025}

\begin{document}
\label{firstpage}
\pagerange{\pageref{firstpage}--\pageref{lastpage}}
\maketitle
% Abstract of the paper (250 words)
\begin{abstract}
 Comet 7P/Pons-Winnecke was observed from the Calar Alto Observatory (Spain) for four months during the 2021 inbound apparition. Broad-band visible images were taken between 1.71 and 1.25~AU pre-perihelion, while long-slit spectrophotometric data were taken at $\sim$~1.25~AU pre-perihelion. This dataset has been complemented with three $r$-Sloan images observed from Zwicky Transient Facility (ZTF) to model the physical properties and loss rate of the dust with a forward Monte Carlo dust tail code. The model fits the observed isophotes well for most observations. The peak dust production rate was measured at 83~kg~s$^{-1}$, 15 days after perihelion. The particle terminal speed ranges from 3~m~s$^{-1}$ for 0.1~m particles to 23~m~s$^{-1}$ for 5~$\mu$m particles. Regarding the gas production from spectra, CN and C$_2$ show asymmetric emission between the sunward and antisunward directions beyond the data uncertainties and error propagation, while a clear asymmetry for C$_3$ cannot be definitively claimed. {\color{black} Average production rates for CN, C$_2$, and C$_3$ near 2021 perihelion are 1.15 $\times 10^{24}$, 2.32$\times 10^{24}$, and 1.69$\times 10^{23}$ s$^{-1}$, respectively.} The dust-to-gas mass ratio value {\color{black} is estimated to be} around 2, suggesting a dust-rich composition. Based on the gas composition and the $Af\rho$ value, we classify 7P/Pons-Winnecke as having a typical composition for Jupiter Family comets, with some C$_3$ depletion. Given the limited previous knowledge, our work contributes to expanding the understanding of the activity and characteristics of 7P/Pons-Winnecke.
\end{abstract}

% Select between one and six entries from the list of approved keywords.
% Don't make up new ones.

%%%%%%%%%%%%%%%%%%%%%%%%%%%%%%%%%%%%%%%%%%%%%%%%%%

%%%%%%%%%%%%%%%%% BODY OF PAPER %%%%%%%%%%%%%%%%%%

\section{Introduction}

Comet 7P/Pons-Winnecke (hereafter 7P) is a short-period comet first discovered by Jean Louis Pons in June 1819 and later rediscovered by Friedrich Theodor Winnecke in 1858. Although it is a well-known comet, there is limited detailed information about it. During its most recent perihelion passage (2021), we observed 7P through imaging and long-slit spectroscopy, which allowed us to study the gas and dust activity of the comet.\par

The current orbital period of 7P is 6.32 years, and has a Tisserand parameter, $T_J \approx 2.68$, calculated using orbital elements from \href{https://ssd.jpl.nasa.gov/tools/sbdb_lookup.html#/?sstr=7P}{Jet Propulsion Laboratory}\footnote{\url{https://ssd.jpl.nasa.gov/tools/sbdb\_lookup.html\#/?sstr=7P}}
. According to the criteria defined by \cite{1997Icar..127...13L} \textcolor{black}{ \citep[based on the previous work of][]{1996ASPC..107..173L}}, 7P belongs to the Jupiter Family. This family is generally accepted to have originated in the Kuiper Belt, and their orbits are significantly influenced by gravitational interactions with Jupiter \citep[see][] {1997Icar..127...13L}.\par

The perihelion distance ($q$) of 7P has undergone significant changes, increasing from $q=0.77$~AU in 1858 to $q=1.26$~AU in 1989. The current perihelion distance is $q=1.23$~AU. According to \citet{Kinoshita}, the perihelion distance is expected to decrease, potentially reaching $q=0.84$~AU during the 2073 apparition.\par
7P has been associated with the 1998 June Boötids meteor shower \citep{2002EM&P...88...27T}. According to these authors, the meteoroids were released by the comet during its 1819 and 1869 perihelion passages. By the mid-1900s, these meteoroids began transitioning into Earth-intersecting orbits, leading to the observed meteor shower in 1998. Furthermore,
 \cite{2012epsc.conf...68M} also reported the observation of meteoroids originated from 7P.\par

The nucleus of 7P is relatively large, with a radius reported as $R_n = 2.6$~km by~\cite{2004come.book..223L} and slightly smaller, $R_n = 2.24$~km by~\cite{2005A&A...444..287S}. Regarding its activity, 7P  was weakly active on its 2021 apparition. \cite{2024SoSyR..58..456N} monitored 7P from $r_h=2.31$~AU to $r_h=1.23$~AU, and reported activity began at $r_h = 1.76$~AU and persisted for approximately 13 months. During this perihelion passage, \cite{2021ATel14486....1K} reported a pre-perihelion outburst on 2021 March 19. Nine additional outbursts were detected post-perihelion, between Jun 5 and Aug 25, 2021, \citep[see][and references therein]{2022PSJ.....3..173L}. \cite{2022PSJ.....3..173L} suggested the large number of outbursts detected in 7P (surpassed only by 29P/Schwassmann–Wachmann 1) may be partly due to favorable observational \textcolor{black}{geometry}. \par
During past apparitions, comet 7P has not exhibited particularly high activity levels. For example, production rates of different comets were measured by \cite{COCHRAN2012144} and \cite{1996ApJ...459..729F}. \cite{COCHRAN2012144} reported upper limits for CN production in 7P of $5.75 \times 10^{23}$~s$^{-1}$ at \( r_h = 1.73 \, \text{AU} \) in the 1989 apparition. In contrast, \cite{1996ApJ...459..729F} estimated the CN production rate when 7P was at \( r_h = 1.42 \, \text{AU} \) obtaining an approximate value of \( 4 \times 10^{24} \, \text{s}^{-1} \) for the same apparition. Comparing both estimates, the latter value appears significantly larger than expected when scaled with the heliocentric distance following a power law of -2.7 \citep{AHEARN1995223}. More recently, \cite{Blain2022Dust} reported CN production rates during the 2021 apparition that were a factor of 2--4 higher than those reported by \cite{COCHRAN2012144}.\par %However, this discrepancy could be attributed to outbursts, which were observed during the 2021 apparition.\par 
During the definition phase of the Comet Interceptor mission (ESA-JAXA), adopted on June 8, 2022, comet 7P was considered a backup target %\footnote{The current mission profile discards 7P as a backup target}
in case a DNC or an interstellar object approaching the inner solar system had an unfavourable orbital configuration for being intercepted by the spacecraft~\citep{Schwamb_2020}. Prompted by that circumstance, and especially by the scarcity of data from 7P, we began tracking its activity from the Calar Alto Observatory from 1.71 to 1.25~AU, as the comet was approaching perihelion. The goal was to characterize the dust environment and gas emissions in the visible. The observationally favorable apparition in 2021 allowed the study of 7P, including its taxonomy, gas emission rates, dust loss mass, and dust-to-gas mass ratio. \textcolor{black}{Given the frequent outbursts of 7P during its 2021 apparition \citep{2022PSJ.....3..173L}, we compared their reported outburst dates with our observations. Only one outburst was reported before June (on March 19), which does not affect our observations (taken until March 18 and after April 13). We did not find a notable influence of the outbursts also reported by the authors on June 5 and 9 in the June 14 ZTF image, perhaps due to the relatively lower intensities of these events. However, we find significant deviations between our model and the observations in the June 4 ZTF image, indicating the influence of the outburst that took place on June 2, according to \cite{2022PSJ.....3..173L}.
}\par 
In Section~\ref{sect:observations}, we present the data. The dust tail is modelled using a Monte Carlo code in Section~\ref{sect:MC_dust}, while the gas profiles are analyzed with the Haser model in Section~\ref{sect:gas_profiles}, \textcolor{black}{and Section~\ref{sect:dust-to-gas} is devoted to estimating the dust-to-gas mass ratio}. Finally, we summarize our findings and provide conclusions in Section~\ref{sect:conclusions}.

\section{Observations and data reduction}\label{sect:observations}
The activity of comet 7P, both from the gas and dust standpoints, was tracked from the Calar Alto Observatory (CAHA, near Almeria, Spain) from January 12 to June 1, 2021. We used the CAFOS instrument mounted on the 2.2m telescope, allowing for imaging and long-slit spectroscopy. 

The images and long-slit spectra were planned to be acquired every two weeks. However, bad weather conditions prevented us from doing a coarse monitoring. Data acquisition, reduction and calibration are described in the following sections. 

\subsection{Imaging photometry} 
The CAFOS instrument, in its imaging mode, provides a FOV of $18' \times 18'$, with a pixel size of 0.53 arcsec.  Appropriate bias and flat field frames were taken each night. \textcolor{black}{Images were acquired through broadband V and R Johnson-Cousins filter. R filter is the most suitable for analyzing the dust activity evolution, as the molecular emissions within the band should be negligible compared to the signal from the dust continuum.} The data reduction procedure followed standard techniques, such as bias subtraction and flat-field correction. The sky contribution could be removed as estimating its value from the edges of the frame since the dust coma does not fill the FOV of the CCD. 
%All images were acquired through broadband R Johnson-Cousins filter, which is the most suitable for analyzing the dust activity evolution, as the molecular emissions within the band should be negligible compared to the signal from the dust continuum. The data reduction procedure followed standard techniques, such as bias subtraction and flat-field correction. The sky contribution could be removed as estimating its value from the edges of the frame since the dust coma does not fill the FOV of the CCD. 
The pixel coordinates of the result frames are shifted to zero at the optocenter. This peak position or optocenter is determined by a two-dimensional (2D) Gaussian fit of the central brightest area of the comet image. The processed and aligned images of individual nights are ‘co-added’, applying median-averaging to improve the signal-to-noise ratio and, at the same time, to remove, or at least reduce, signatures from cosmic rays and from background object trails due to the telescope tracking during the cometary exposures.

We completed the CAFOS image set by including three $r$-Sloan images downloaded from the Zwicky Transient Facility (ZTF) obtained between May 23 and June 14, 2021. The ZTF images were already reduced, and a zero-point was provided for absolute calibration.

\begin{figure}
\includegraphics[angle=0,
trim={6cm 3cm 8cm 2cm},clip,
width=\columnwidth]{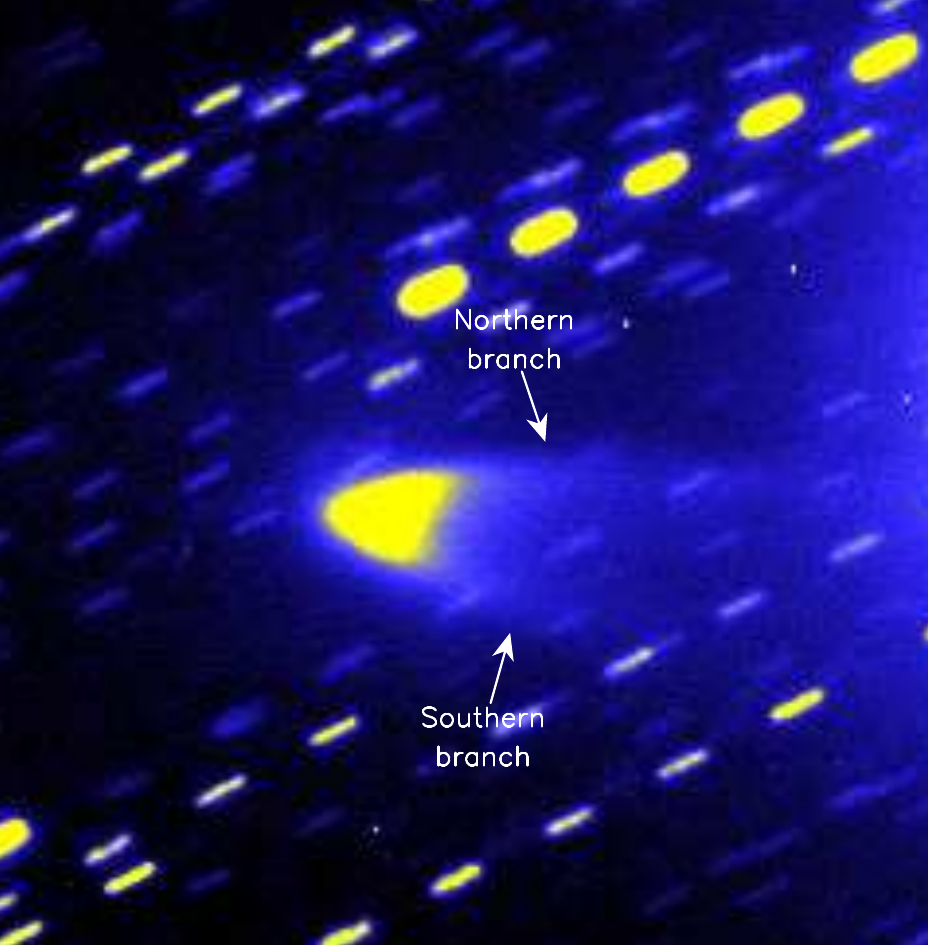}   
\caption{Image of 7P obtained with CAFOS at the 2.2m telescope at Calar Alto Observatory on May 15, 2021, see Table \ref{tab:logobs}, code (h), showing the diffuse northern and southern branches. The image dimensions are 48521$\times$58224 km projected on the sky at the comet distance. North is up, East to the left. The position angle of the Sun-to-comet radius vector is 250.9$^\circ$.
\label{fig:CAHAimage}}
\end{figure}

The geometric parameters of the observations for all the images, CAHA and ZTF,  are shown in Table~\ref{tab:logobs}. The comet tail is typically characterized by a fan-like appearance with two or more (generally two) diffuse jet-like northern and southern features directed in the anti-sunward direction, as was reported in earlier apparitions of 7P \citep[see][and references therein]{1989AJ.....98.2322S}. These features are indicated in the CAFOS image of Fig.~\ref{fig:CAHAimage}.  

%\begin{figure}
%\includegraphics[angle=0,trim={5cm 9cm 5cm 10cm},clip,width=\columnwidth]{ZTF_MAY23.pdf}   \caption{Image of 7P/Pons-Winnecke downloaded from the Zwicky Transient Facility, obtained with the 48-inch Palomar telescope on May 23, 2021, see Table \ref{tab:logobs}, code (i).  The image spans approximately 164000$\times$172000 km projected on the sky at the comet distance. North is up, East to the left. The position angle of the Sun-to-comet radius vector is 250.3$^\circ$.
%\label{fig:ZTFimage}}
%\end{figure}

In addition, to better constrain the dust tail model, we used R-band magnitude measurements from the amateur astronomical association \texttt{Cometas\_Obs}, and magnitude measurements extracted from the IAU Minor Planet Center for this apparition from early December 2020 to mid December 2021. The magnitudes from \texttt{Cometas\_Obs} were all taken using known apertures of size projected in the sky in the 1700 to 4800 km range. To refer all images and magnitude measurements to the same brightness level, all the image data were calibrated so that the calculated magnitudes on the images matched those obtained by the \texttt{Cometas\_Obs} association (as shown below). 
%Fig.~\ref{fig:magnitudes} displays the photometric data as a function of time to comet perihelion.

\begin{table*}
  \centering
  \caption{Log of the observations: images. $\Delta_{tp}$
  indicates days since perihelion, $r_h$ is the heliocentric distance, 
$\Delta$ is the geocentric distance, PsAng is the position angle of the extended Sun-to-comet radius vector, 
PsAMV is the negative of the comet heliocentric velocity vector, and PhAng is the phase angle. Nights on which spectra were taken and used for the analysis in this work are indicated with text in bold in the first column.}   
  \label{tab:logobs}
  \begin{tabular}{|l|c|c|r|c|c|c|c|c|c|}
    \hline
    Telescope & Code & Time (UT) & $\Delta_{tp}$ & r$_h$ (au)& $\Delta$ (au) & PsAng($^\circ$) & PsAMV($^\circ$) & PlAng($^\circ$) & PhAng($^\circ$) \\
    \hline
 CAHA 2.2m & (a) &2021-Feb-15 06:07:15   &   --100.9 & 1.7096 &  1.3259 & 282.174  & 302.727 &  --13.20000  &  35.1923  \\
 CAHA 2.2m & (b) &  2021-Mar-16 03:30:45   &   --72.0 &1.5067 &  0.9721 &   270.133  & 304.328 &  --22.78589  &  40.5611  \\
 CAHA 2.2m & (c) &  2021-Mar-18 02:44:59   &   --70.0 &1.4939 &  0.9510 &   269.294  & 304.402 &  --23.43368  &  40.9678 \\
 CAHA 2.2m & (d) &  2021-Apr-13 02:34:10   &   --44.0&1.3463 &  0.7082 &   258.939  & 304.522 &  -30.57247  &  46.6635 \\
 CAHA 2.2m & (e) &  2021-May-04 02:33:37   &   --23.0 &1.2664 &  0.5613 &   252.799  & 302.248 &  --31.56349  &  50.6171 \\
\textbf{CAHA 2.2m} & (f) &  2021-May-08 02:08:13   &   --19.0 &1.2563 &  0.5391 &   251.990  & 301.332 &  -30.88760  &  51.1173 \\
 \textbf{CAHA 2.2m} & (g) &  2021-May-13 01:56:55   &   --14.0 &1.2463 &  0.5140 &   251.173  & 299.835 &  -29.55418  &  51.5644 \\
 \textbf{CAHA 2.2m} & (h) &  2021-May-15 01:55:16   &   --12.0 &1.2431 &  0.5049 &   250.911  & 299.104 &  -28.86219  &  51.6803 \\
 ZTF Palomar 48-inch & (i)  & 2021-May-23 11:06:26   & --3.7 &  1.2350 &  0.4731 &   250.259  & 294.994 &  -24.97011  &  51.7267 \\
 ZTF Palomar 48-inch & (j) & 2021-Jun-04 09:36:10   &  +8.3 & 1.2384 &  0.4467 &   250.725  & 285.174 &  -16.98500  &  50.4880 \\
 ZTF Palomar 48-inch & (k) & 2021-Jun-14 09:33:24   &   +18.3 & 1.2547 &  0.4422 &   252.576  & 272.501 &   -8.89358  &  48.3132 \\ \hline
  \end{tabular}  
  \end{table*}

\subsection{Long-slit spectrophotometry}
\begin{figure}
\includegraphics[angle=0,
width=\columnwidth]{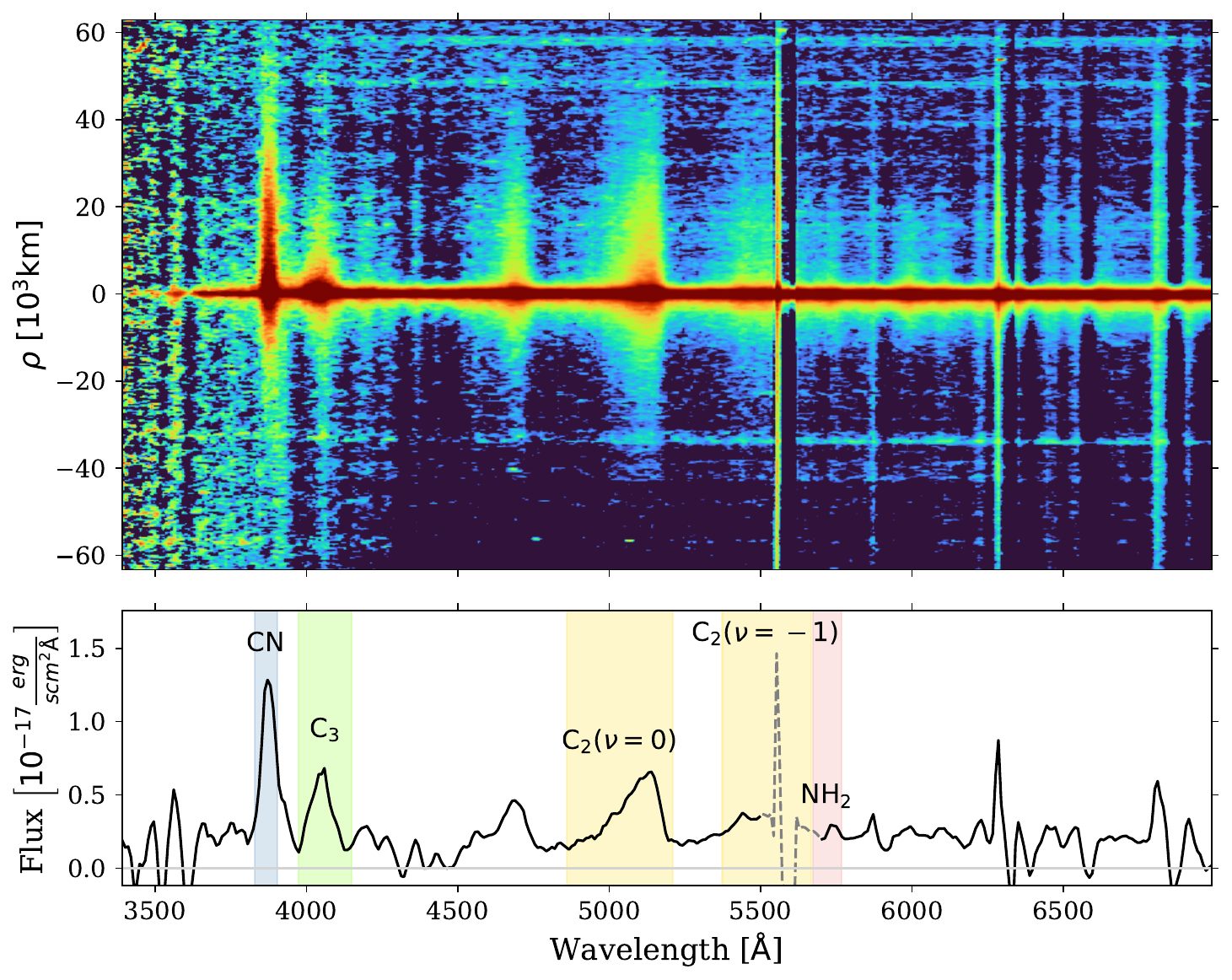}   
\caption{\textcolor{black}{The top panel shows the stacked flux-calibrated spectrum of comet 7P acquired on May 13, 2021, centered on the nucleus. Assuming the nucleus is placed at $\rho = 0$ km, the --y axis refers to the anstisun direction, and +y axis to the Sun direction. The spatial scale is 158 km/pix and the spectral resolution is 0.94 nm/pix. The colour code stretches from 0 to $3.0\times 10^{-17}$ $\rm ergs~cm^{-2} s^{-1} A^{-1}$. The bottom panel shows the averaged flux along the spatial axis to highlight emission bands. A sky background subtraction issue is visible between the C$_2$($\nu = 0$) and NH$_2$ bands (dashed grey line). The main cometary emission bands (CN, C$_2$, C$_3$, and NH$_2$) are highlighted. The slit was positioned at a position angle (PA) of 70°, covering spatial profiles in approximately the Sun-antisun direction as projected on the plane of the sky.}
\label{fig:CAHAspectrum}}
\end{figure}

Spectrophotometric observations were taken on March 18, May 8, 13, 14, 15 and June 1, 2021. Sky conditions prevented us from retrieving scientific information on the nights of March 18, May 14 and June 1. Furthermore we were unable to derive NH$_2$ column density profiles on the remaining nights (May 8, 13, and 15) its emission band overlapped with a bright sky region. We used the grism B-400 (see \href{https://www.caha.es/CAHA/Instruments/CAFOS/cafos22.html}{CAFOS})
and a slit width and length of 2.5'' and 9', respectively. This resulted in an observable spectral range between 340 and 740 nm with a wavelength scale of $\approx 0.9$ nm per pixel and a spatial scale of 0.53 arcsec per pixel. For every night, the slit of the spectrograph was oriented at a PA of 70$^\circ$ which lay parallel to the Sun–to-comet line projected on the plane. Three spectra were acquired with this configuration each night. For absolute calibration, we obtained observations of the spectrophotometric standard star BD+33D2642 with a slit width of 10'', the same slit length, and an airmass similar to that of the comet observations. Every spectra was bias subtracted, flat fielded, wavelength calibrated (using He–Ar reference spectra), extinction corrected (using the standard extinction curve for Calar Alto observatory), and finally flux calibrated. The cometary emission did not fill the entire slit and the sky flux could be well determined from the sky retrieved at the edges of the frames. Fig.~\ref{fig:CAHAspectrum} shows the flux-calibrated spectrum of 7P as observed on May 13, 2021. \textcolor{black}{The entire CCD extends from the optocenter up to $\rho =$  80,000 km in both directions. In the top panel, the anisotropic emission of the gas profiles can be directly identified by comparing between sunward and antisunward directions. The lower panel shows the 1D spectrum, averaging both directions.}

\section{Monte Carlo dust tail modelling}\label{sect:MC_dust}

\begin{figure}
\includegraphics[angle=0,
trim={2cm 2cm 5cm 2cm},clip,
width=\columnwidth]{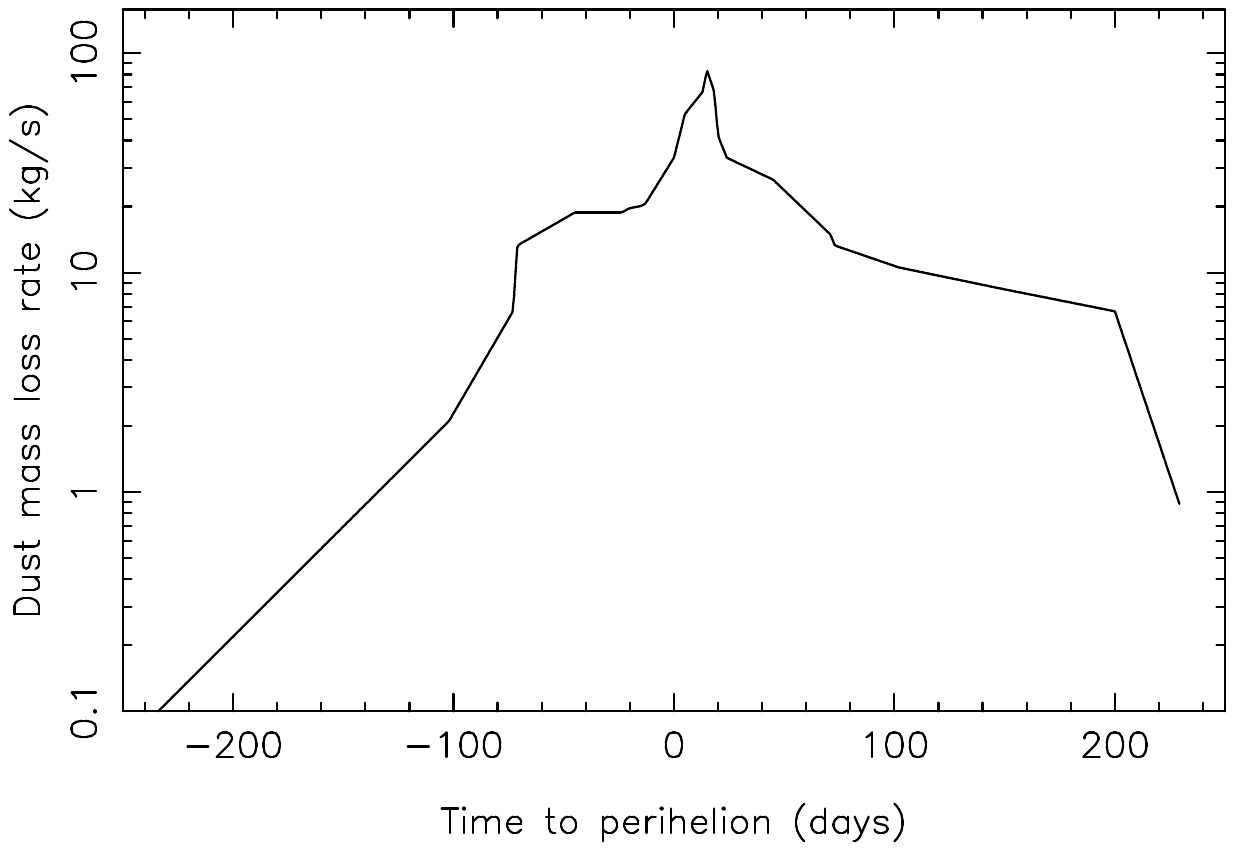}   
\caption{The derived dust mass loss rate as a function of time to perihelion for comet 7P.   
\label{fig:masslossrate}}
\end{figure}

\begin{table}
    \centering
    \caption{Physical parameters for the best-fit dust ejection model.}
    \label{tab:parameters}
    \begin{tabular}{|l|l|l|}
    \hline
    Parameter & Value & Assumed (ref)/ \\ 
               &  & Derived \\ \hline
    Particle geometric albedo & 0.04  & Assumed \\
    Phase function  &  & D. Schleicher \\
    Minimum particle radius & 5 $\mu$m  & Assumed \\
    Maximum particle radius  & 0.1 m  & Assumed \\
    Particle density  & 1000 kg m$^{-3}$) & Assumed \\
    Size distribution power index & --3.7 & Assumed \\
Minimum speed ($r$= 0.1m)& 3 m s$^{-1}$ & Derived\\ 
Maximum speed ($r$=5 $\mu$m) & 
23 m s$^{-1}$ & Derived   \\
    Peak production rate & $\sim$100 kg s$^{-1}$ & Derived \\
    Time of peak ejection & 15 d post-perihelion & Derived \\ 
Nucleus rotational parameters & $I$=58$^\circ$, $\Phi$=78$^\circ$ & 
\cite{1989AJ.....98.2322S} \\
Latitude box of active region & [+30$^\circ$,+60$^\circ$] & Derived \\
     \hline
     \end{tabular}
  \end{table}

\begin{figure*}
\includegraphics[angle=0,trim={0 6cm 4cm 2.5cm},clip,
width=\textwidth]{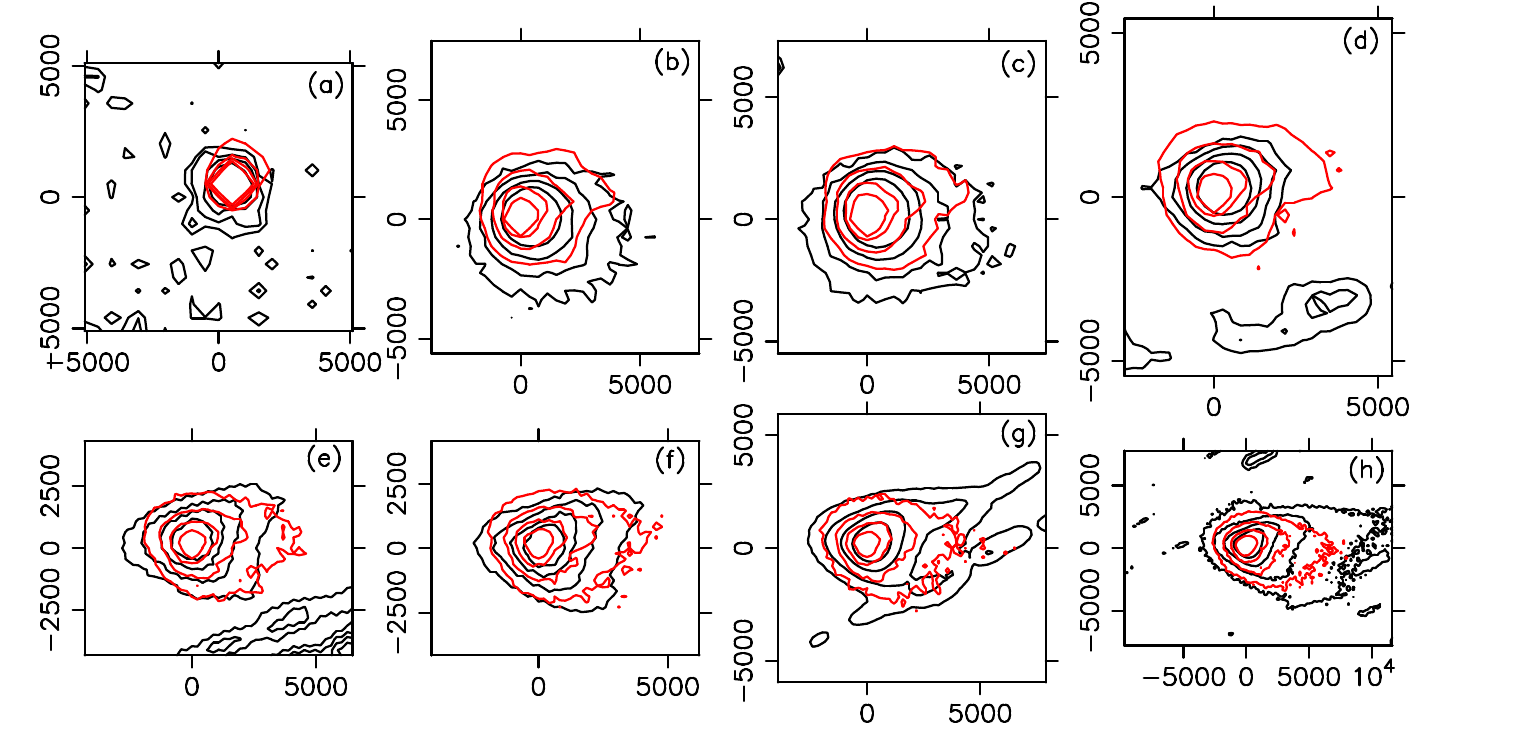}   
\caption{Black contours are the observed isophotes for the images obtained with CAFOS at the CAHA 2.2m, and red contours are the model isophotes. Labels (a) through (h) refer to Table \ref{tab:logobs}. The innermost isophote levels are 1.6$\times$10$^{-13}$ for images labelled as (a), (b), and (c), and 6.4$\times$10$^{-13}$ for images (d) through (h), in solar disk intensity units. Isophotes decrease in factors of 2 outwards.
X- and Y-axes are labelled in projected distances from the comet nucleus in km. North is up East to the left. 
\label{fig:CAHA_iso}}
\end{figure*}

\begin{figure}
\includegraphics[angle=0,trim={0 12cm 9cm 3.5cm},clip,width=0.99\columnwidth]{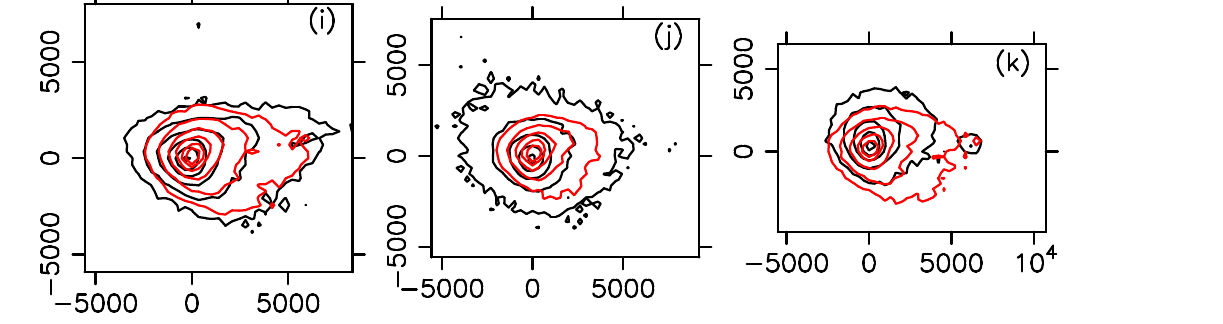}   
\caption{Black contours are the observed isophotes for the ZTF images obtained at the 48-inch Palomar telescope and red contours are the model isophotes. Labels (i), (j), and (k) refer to Table \ref{tab:logobs}. The innermost isophote levels is 3.2$\times$10$^{-12}$ solar disk intensity units for the three images. Isophotes decrease in factors of 2 outwards. X- and Y-axes are labeled in projected distances from the comet nucleus in km. North is up, East to the left. 
\label{fig:ZTF_iso}}
\end{figure}

\begin{figure}
\includegraphics[angle=0,
width=0.5\textwidth]{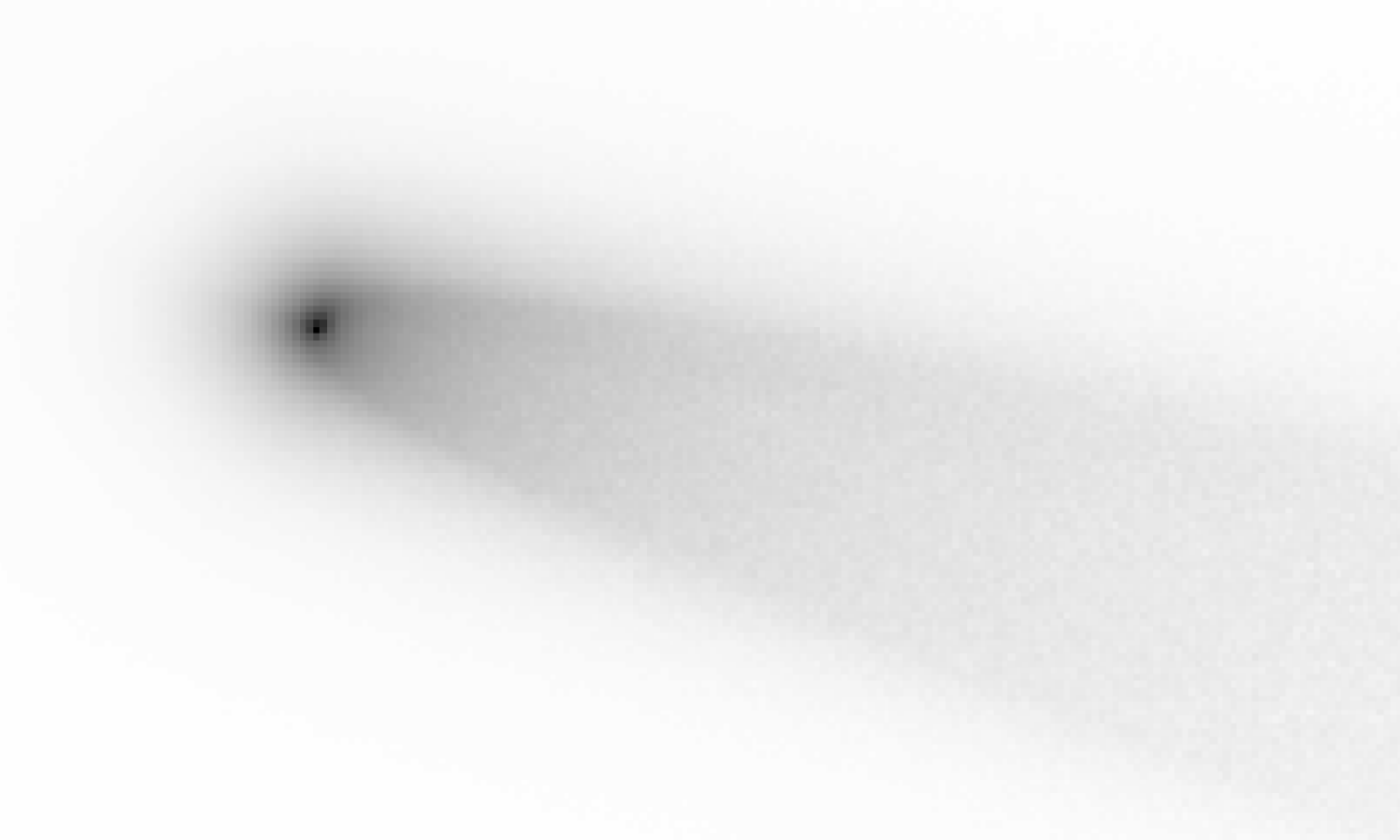}   
\caption{Simulated ZTF image for 2021 June 14 (see label (k) in Table \ref{tab:logobs}), showing the northern and southern diffuse branches. Compare this with the similar appearance of the ZTF image on the lowermost right panel of Fig.~\ref{fig:outburst}. North is up East to the left. 
\label{fig:ZTF_simul}}
\end{figure}

\begin{figure}
\includegraphics[angle=0,trim={4cm 2cm 4cm 1cm},clip,width=0.99\columnwidth]{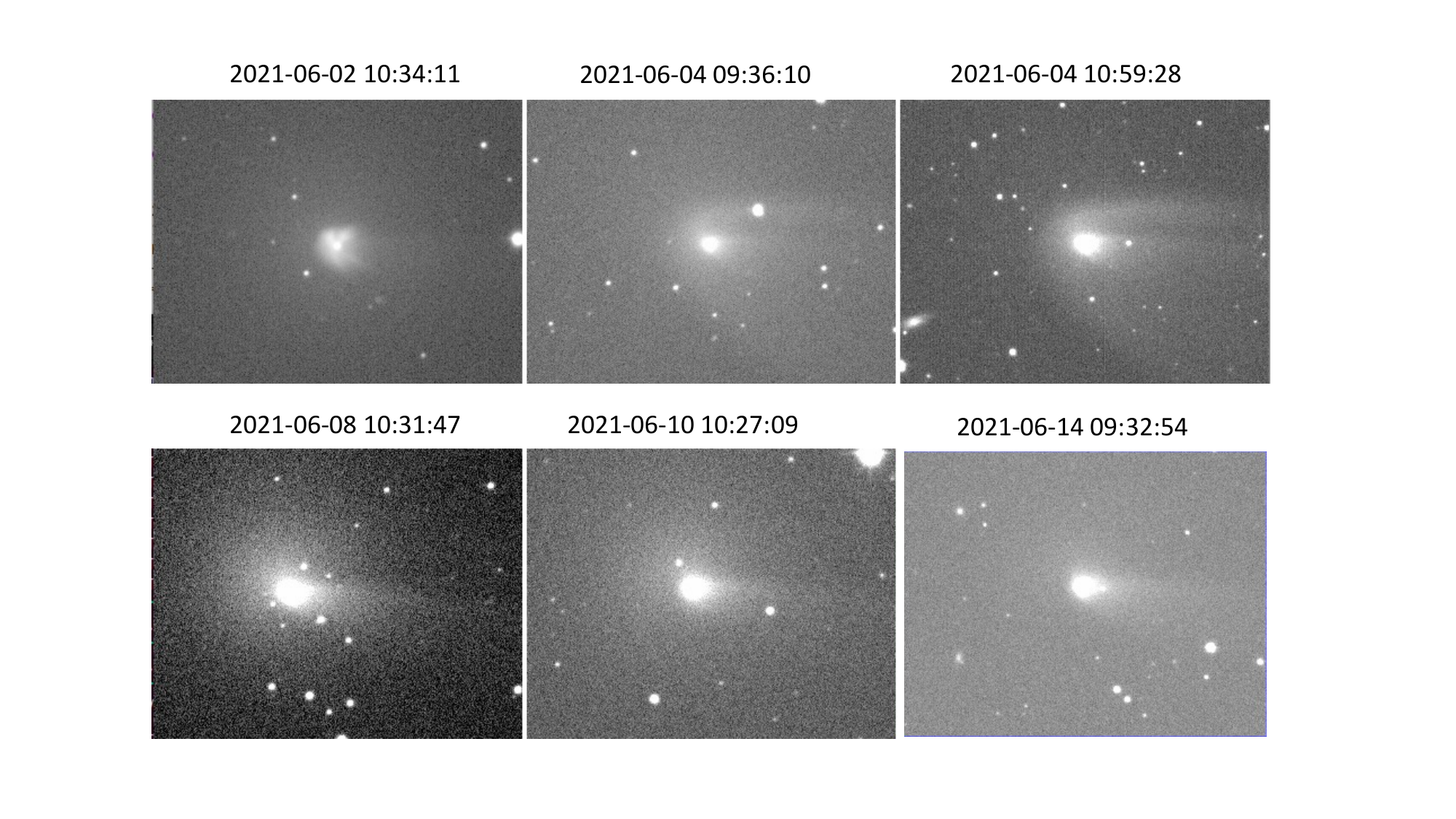}   
\caption{Sequence of ZTF images showing the occurrence of an outburst starting around 2021 June 2, with ejection of fine dust particles showing an external dust shell on 2021 June 4, and returning back to baseline activity on 2021 June 14. The observing date (UT) is at the top of each panel. North is up East to the left.
\label{fig:outburst}}
\end{figure}

\begin{figure}
\includegraphics[angle=0,
trim={1cm 0.5cm 3cm 1cm},clip,width=\columnwidth]{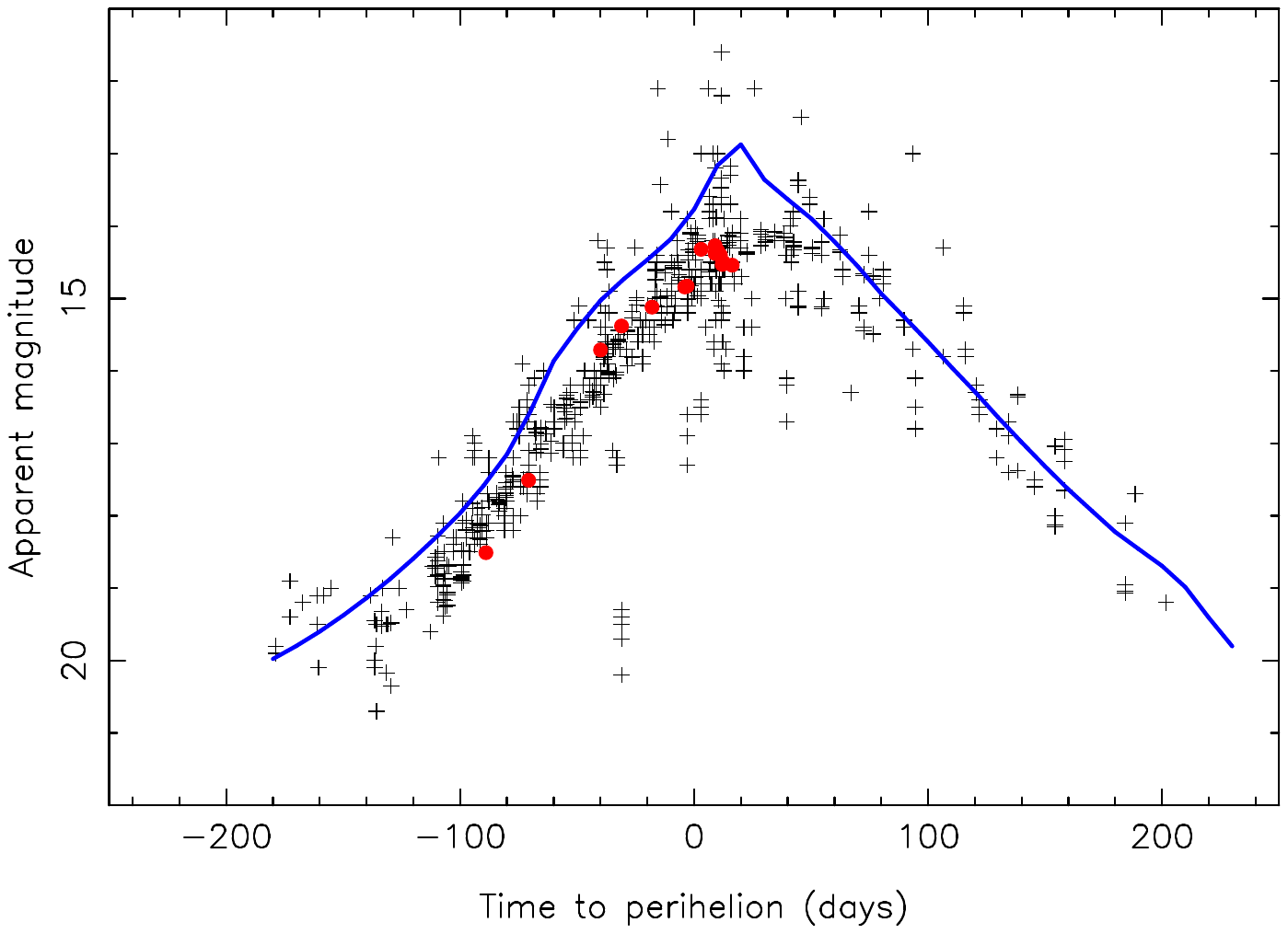}   
\caption{Apparent magnitudes vs. time to 7P perihelion. The black crosses are the "nuclear" magnitudes extracted from the Minor Planet Center database. The filled red circles are the R-band magnitudes from \texttt{Cometas\_Obs} association.  The blue thick line corresponds to the modeled R-band magnitudes calculated for the whole time span from about --200 to +200 days since perihelion.   
\label{fig:magnitudes}}
\end{figure}

To determine the physical properties and dust loss rate, we employed our forward Monte Carlo dust tail code \citep[see][and references therein]{2022Univ....8..366M}.  The code is specifically designed for the outer coma and tail regions, where the gravitational influence of the nucleus and the gas drag forces on the dust particles are negligible. Therefore, the initial particle speeds actually represent the terminal speeds at distances of approximately 20 nuclear radius. The particle motion is then influenced solely by solar gravity and radiation pressure. The particles are assumed as compact spheres so that the radiation pressure is directed radially away from the Sun, as does the gravity (in the opposite sense), so that they describe a Keplerian trajectory around the Sun.

The code calculates the orbits of particles based on their initial velocities and the $\beta$ parameter, the ratio of the radiation pressure force to the gravity force, expressed as $\beta = \frac{C_{pr} Q_{pr}}{2\rho r}$ \citep{1968ApJ...154..327F}, where $C_{pr} = 1.19 \times 10^{3}$ kg m$^{-2}$ is the radiation pressure coefficient, $Q_{pr}$ is the scattering efficiency for radiation pressure \citep[supposed to be $Q_{pr}$=1,][]{1979Icar...40....1B}, $\rho$ is the particle density (assumed at 1000 kg m$^{-3}$), and $r$ is the particle radius. The heliocentric position of each particle is calculated for the observation date and projected onto the sky plane by calculating their $(N,M)$ coordinates 
\citep[see][]{1968ApJ...154..327F}. However, we also keep track of the third coordinate $L$ (the axis directed from the nucleus toward the observer, normal to $N$ and $M$), needed for the computation of the nucleus brightness as shown below.

In the Monte Carlo simulation, a large number of particles ($\gtrsim 10^7$) are tracked, with their brightness depending on size and geometric albedo, taken as $p_v = 0.04$, a typical value for comet dust. To correct for the phase function effect, we used the phase function constructed by David Schleicher \footnote{\url{https://asteroid.lowell.edu/comet/dustphase/details}} from observations of several comets. The final synthetic brightness images are generated by adding up the contributions from all sampled particles, depending on the mass loss rate, the assumed size distribution, and the particle velocities. The size distribution follows a power-law function, bounded by a minimum and a maximum radius  ($r_{min}$ and $r_{max}$) and a specific power exponent, $\kappa$, i.e., $n(r) \propto \int_{r_{min}}^{r_{max}} r^\kappa$. Alongside the dust brightness, we also accounted for the nucleus brightness. The nucleus radius is taken from the value reported by \cite{2005A&A...444..287S}, i.e., $R_n$=2.24 km, for a geometric albedo of 0.04. To correct the phase effect on the nucleus brightness, we assumed a linear phase correction with a coefficient of 0.03 mag deg$^{-1}$.  The actual nucleus brightness is reduced by the optical depth of the coma in the line of sight toward the observer, which is computed as the summation extended to all particles falling on the pixel corresponding to the photocenter of the image, along the line of sight between the nucleus and the observer (i.e., for those particles having $L$>0), integrated over time of the quantity:
\begin{equation}
    \tau(r) = \frac{3 Q_{ext} M}{4 r \rho}, 
\label{eq:opdepth}
\end{equation}
where $Q_{ext}$ is the extinction coefficient, and $M$ is the integrated mass column (kg m$^{-2}$). We take $Q_{ext}=2$ (the Fraunhoffer approximation), a value appropriate for particles in the $r \gtrsim$ 1 $\mu$m domain for incident red wavelengths.   

%\begin{figure}
%\includegraphics[angle=0,
%trim={1cm 0.5cm 3cm 1cm},clip,width=\columnwidth]{magnitudes_comparison.pdf}   
%\caption{Apparent magnitudes vs. time to 7P perihelion. The black crosses are the "nuclear" magnitudes extracted from the Minor Planet Center database. The filled red circles are the R-band magnitudes from \texttt{Cometas\_Obs} association.  The blue thick line corresponds to the modeled R-band magnitudes calculated for the whole time span from about --200 to +200 days since perihelion.   
%\label{fig:magnitudes}}
%\end{figure}

To find the best possible fit for the observed tails, we followed a trial-and-error procedure. Given the large number of free parameters of the model, the solution cannot unfortunately be unique. We rely on previous typical values of physical parameters found for other comets to obtain a set of reasonable values to start the procedure. Thus, in addition to the parameters already described, we set the particle density to 1000 kg m$^{-3}$, and we initialize the power-law index to 
$\kappa$=--3.5, and the limiting radii of the size distribution to $r_{min}$=1 $\mu$m and $r_{max}$=1 cm. The velocity was parameterized by a customary function of $\beta$, heliocentric distance, and solar zenith angle as $v = v_0 {\beta}^\gamma \cos(z)^n / \sqrt{r_h} $ km s$^{-1}$, where $r_h$ is expressed in au, $v_0$ (km s$^{-1}$), and $\gamma$, are free parameters, $z$ is the solar zenith angle at the emission point, and $n$=0.5, which constitutes the approximate dependence of terminal speeds on the cosine of zenith angle found by  \cite{1997Icar..127..319C}. The dust mass loss rate, $\frac{dM}{dt}(t)$,  was set to a Gaussian function with peak production rate occurring sometime relative to perihelion, $t_0$, and a certain full-width at half-maximum (FWHM). After some experimentation with the code, it was clearly apparent that the dust loss rate profile could not be modelled as a Gaussian function and, more importantly, that an isotropic emission pattern was unable to reproduce the observed tail structure seen in the images (see Fig.~\ref{fig:CAHAimage}). Instead, following \cite{1989AJ.....98.2322S}, we considered a rotating nucleus with an active area on it. The rotational parameters of the nucleus are the obliquity, $I$, which is the angle subtended by the rotational axis and the normal to the comet's orbit plane, and $\Phi$,  the argument of the subsolar meridian at perihelion  \citep[see]{1981AREPS...9..113S}. We adopted the rotational parameters derived by 
\cite{1989AJ.....98.2322S}, i.e., 
$I$=58$^\circ$, and $\Phi$=78$^\circ$. The active area location was placed to the latitude region of 45$\pm$15$^\circ$. \textcolor{black}{Initially, we assumed that emission originated solely from the active area, which gave overall better fits to the ZTF images. However, the observed isophotes were not well reproduced in all cases, particularly in capturing the triangular shape observed in CAHA images d-h. To achieve overall agreement with the observed isophotes, we found it necessary to include background emission from the entire illuminated hemisphere. We tested various percentages of background emission and determined that the best overall agreement was achieved when 80\% of the total particle emission came from the background and 20\% from the active area. In particular, we prioritized obtaining a better fit with the CAHA images, as they are not affected by the outbursts.} In all cases, the emission of particles followed the radial direction from the emission point outwards. \par

The resulting particle velocity function was $v$=0.04$\beta^{0.2} \cos(z)^{0.5} / \sqrt{r_h} $ km s$^{-1}$, which gives maximum speeds ($z$=0$^\circ$) ranging from 23 to 3 m s$^{-1}$ for particles in the 5 $\mu$m to 0.1 m radius range, which is the final domain of particle sizes considered, with a power-law distribution of index --3.7. For comparison, \cite{2024SoSyR..58..456N} obtained mean particle speeds of 15 m s$^{-1}$ for this comet.

The presence of particles as large as 0.1 m is justified as it has been detected meteoroids in the Earth's atmosphere from 7P as massive as 260 kg \citep{2012epsc.conf...68M}, which for the supposed density of 1000 kg m$^{-3}$ it would correspond to a sphere of $\sim$0.4 m in radius. Large particles (mm-sized) have also been found by \cite{2024SoSyR..58..456N} to dominate the 7P coma. In line with this, decimeter-sized particles have been observed being ejected from the nucleus of 67P/Churyumov-Gerasimeko \citep{2024A&A...685A.136P}.

Finally, the dust loss rate profile that best fits the observations is given in Fig.~\ref{fig:masslossrate}. This profile shows an asymmetrical behaviour with respect to perihelion date, with a maximum of 83 kg s$^{-1}$ some 15 days after perihelion \textcolor{black}{(in agreement with \cite{2022PSJ.....3..173L}; see their Fig.~5)}, and a sustained activity level several weeks before perihelion. This asymmetrical behaviour, with higher brightness levels after perihelion passage than pre-perihelion, has been found in some other short-period comets such as 6P/d'Arrest, 30P/Reinmuth, or 104P/Kowal \citep[see][]{2011SoSyR..45..330B}. \textcolor{black}{In a survey of JFCs, \cite{2013Icar..225..475K} also found that, in most cases, activity was higher post-perihelion.} The high obliquity might be responsible for that behaviour \citep{1981AREPS...9..113S,2019A&A...623A.120M}, although it is difficult to reveal the nature of cometary outgassing from the heliocentric water production rates \citep{2019A&A...623A.120M}.  Concerning the maximum dust loss rate, \cite{2024SoSyR..58..456N} estimated 140 k s$^{-1}$ near perihelion, which is higher than but not too far from our estimate taken into account the many uncertainties in both approaches.

A summary of the relevant model parameters is given in Table \ref{tab:parameters}, and the resulting fits to the CAHA and ZTF images are given in Fig.~\ref{fig:CAHA_iso} and Fig.~\ref{fig:ZTF_iso}. 
A representative image simulation of the ZTF image on June 14 (label k in Table \ref{tab:logobs}) is given in Fig.~\ref{fig:ZTF_simul}, where the relevant features depicted in Fig.~\ref{fig:CAHAimage}, i.e., the very diffuse northern and southern branches, are seen.  As it is seen from the isophote maps, the model isophotes fit reasonably well the observed isophotes for the whole four months of observations. A remarkable exception is, however, seen in Fig.~\ref{fig:ZTF_iso}, central panel (j), which shows a clear disagreement in the most external isophote with that of the model. An examination of the sequence of the ZTF images in the time interval from 2021 June 2 to 2021 June 14 shows the occurrence of an outburst (see Fig.~\ref{fig:outburst}), that took place on or around 2021 June 2\textcolor{black}{. This outburst is characterized by the ejection of very small particles that form a short lasting envelope around the dust coma} superimposed on the previously described features on the 2021 June 4 image. This outburst, which was not modelled, was already reported by \cite{2024SoSyR..58..456N}. Comet 7P has experienced several outbursts of activity during this and previous apparitions \citep[see][]{2021ATel14486....1K,2022ATel15772....1K,
2024SoSyR..58..456N}. \textcolor{black}{Thus, the majority of the data fitted is unaffected by the presence of outbursts, we can state that the model accurately reflects the underlying activity}.

In addition to the image fitting, we also run the model to compute the synthetic apparent magnitudes in the time interval from 180 days pre-perihelion to 230 days post-perihelion, in time steps of 5 days, using the same apertures as those used by the amateur association \texttt{Cometas\_Obs} (see previous section). In Fig.~\ref{fig:magnitudes} we show the results of the modelling (thick blue line), that reproduces reasonably well the photometric data in such a long time interval, in this way complementing the results of the image fitting and providing more support to the model results. \textcolor{black}{However, we note that the synthetic magnitudes slightly overestimate the observed apparent magnitudes mainly in the pre-perihelion branch. This small discrepancy is most likely due to the fact that we have prioritized the whole isophote field observed in the images versus the photometric data.}

\section{Gas production rates}\label{sect:gas_profiles}

\begin{table*}
\centering
\caption{Emission and continuum band extracted from the spectra, $g$-factors and both (parent and daughter) scale lengths \citep[see][and references therein]{AHEARN1995223} used to derive the production rates. We used $g$-factors for C$_2$ and C$_3$ available in \href{https://asteroid.lowell.edu/comet/gfactor}{Lowell Minor Planet Services}$^{\dag}$for every night taking into account the comet heliocentric distance and velocity. $\dag:$ \url{https://asteroid.lowell.edu/comet/gfactor}}
\label{tab:spectral_regions}
\begin{tabular}{ccccccc} \hline
Species & Spectral region &Left-hand cont. & Right-hand cont. & $g$-factor & $l_p$ & $l_d$  \\
& (\text{\AA}) & (\text{\AA}) & (\text{\AA}) & (10$^{-13}$ erg s$^{-1}$ mol$^{-1}$) & (10$^4$ km)  &  (10$^4$ km)   \\ \hline
CN                       & 3830–3905  & 3770–3815 & 3910–3970 & 3.40 // 3.19 // 3.08 &  1.3  & 21.0\\
C$_3$                    & 3975–4150  & 3910–3970 & 3975–4150 & 10                   &  0.3  & 2.7\\
C$_2$ ($\Delta \nu$ = 0) & 4860–5210  & 4780–4850 & 5220–5250 & 4.5                  &  2.2  & 6.6\\
%C$_2$ ($\Delta \nu$ = -1)& 5373–5667  & 5220–5250 & 5769–5790 & 2.05                 &  2.2  & 6.6\\
\hline    
\end{tabular}
\end{table*}

\begin{table*}
\renewcommand{\arraystretch}{1.5} % Ajuste del espaciado entre filas
\centering
\caption{\textcolor{black}{Gas production rates for different species (in units of $10^{23}$ s$^{-1}$), and the logarithmic ratios of the production rates of C$_3$ and C$_2$ relative to CN, denoted as C$_3$/CN and C$_2$/CN, respectively.}
}
\label{tab:production_rates}
\begin{tabular}{ccccccc} % Ajuste de columnas
\hline
Time (UT)            & Direction & $Q_{\text{CN}}$        & $Q_{\text{C}_3}$      & $Q_{\text{C}_2} $      & C$_3$/CN  & C$_2$/CN   \\ \hline 
2021-May-08 03:25:28 & Sun     & $9.38\pm 0.17$  & $1.27\pm 0.03$ & $22.39\pm 0.38$ & -0.87 $\pm$ 0.08 & 0.38 $\pm$ 0.07 \\
                                         & Antisun & $7.78\pm 0.18$  & $1.18\pm 0.03$ & $18.73\pm 0.43$ & -0.82 $\pm$ 0.10 & 0.38 $\pm$ 0.10 \\   \hline
2021-May-13 03:11:12                     & Sun     & $15.57\pm 0.26$ & $1.87\pm 0.03$ & $28.38\pm 0.47$ & -0.92 $\pm$ 0.07 & 0.26 $\pm$ 0.07 \\
                                         & Antisun & $12.77\pm 0.31$ & $1.72\pm 0.04$ & $23.78\pm 0.48$ & -0.87 $\pm$ 0.10 & 0.27 $\pm$ 0.09 \\   \hline
2021-May-15 03:18:06                     & Sun     & $13.12\pm 0.23$ & $2.24\pm 0.04$ & $24.98\pm 0.42$ & -0.77 $\pm$ 0.08 & 0.28 $\pm$ 0.07 \\
                                         & Antisun & $10.50\pm 0.30$ & $1.88\pm 0.07$ & $21.13\pm 0.47$ & -0.75 $\pm$ 0.14 & 0.30 $\pm$ 0.11\\   \hline
\end{tabular}
\end{table*}

We derived CN, C$_2$, and C$_3$ column density profiles ($N$) from spectra taken on May 8, 13 and 15, in the directions probed by the slit as projected on the plane of the sky. To enhance the signal-to-noise ratio, we combine the three individual spectra taken with the same slit configuration each night. Using these stacked spectra, we subtract the underlying dust continuum in each gas emission band by interpolating between the left and right continuum bands  (Table~\ref{tab:spectral_regions}). After isolating the pure gas flux profiles of CN, C$_2$ ($\nu=0$), and C$_3$, we convert these profiles into column densities using the corresponding $g$-factors (Table~\ref{tab:spectral_regions}) with an estimated error below 10\%.\par

The aforementioned column densities profiles were fitted with the Haser model \citep{1957BSRSL..43..740H}. Profiles for both sunward and antisunward directions are analyzed separately. The outflow velocity of the gas adopted in this study is  $v =$~1~km~s$^{-1}$, independent of heliocentric distance. This assumption allows us to calculate $Q / v$, enabling easy comparison with similar studies for active comets (e.g.~\citealt{2024MNRAS.534.1816F}). The parent and daughter scale lengths ($l_p$, $l_d$) are taken from \cite{AHEARN1995223} (see Table~\ref{tab:spectral_regions}) and it is assumed that all scale lengths vary as $r_h^2$. 
\textcolor{black}{To calculate the production rates and their associated uncertainties, we proceed as follows: 
\begin{enumerate}
    \item We prepare the observational data for analysis by binning the observed $\log(N)$ profiles on a cometocentric logarithmic scale, as in~\citet{2021MNRAS.508.1719G}. To estimate the uncertainty for each bin, we consider two values: 10\% of the mean value and the mean absolute deviation of the data within the bin. The final uncertainty is chosen as the larger of these two values.
    \item Thereafter, we simulate 10,000 profiles. Each simulation is created by taking the observed-binned values and adding random variations within the measurement error.
    \item Next, we generate a set of theoretical column density profiles using the activity module from the Python \texttt{sbpy} package. We vary the parameter Q to produce different profiles of N as a function of the projected cometocentric distance $\rho$. 
    \item Afterwards, we compare theoretical and simulated profiles, obtaining the Q value that minimizes the likelihood for each simulated profile. This results in a set of 10,000 optimal Q values, all corresponding to a single observational profile.
    \item Finally, we determine the final Q as the median and the uncertainty as three times the standard deviation of the previous set of optimal Q values.
\end{enumerate}
We apply this procedure for the sunward and the antisunward directions of each spectrum.
}
\par
\begin{figure}
    \centering
    \includegraphics[width=1\columnwidth]{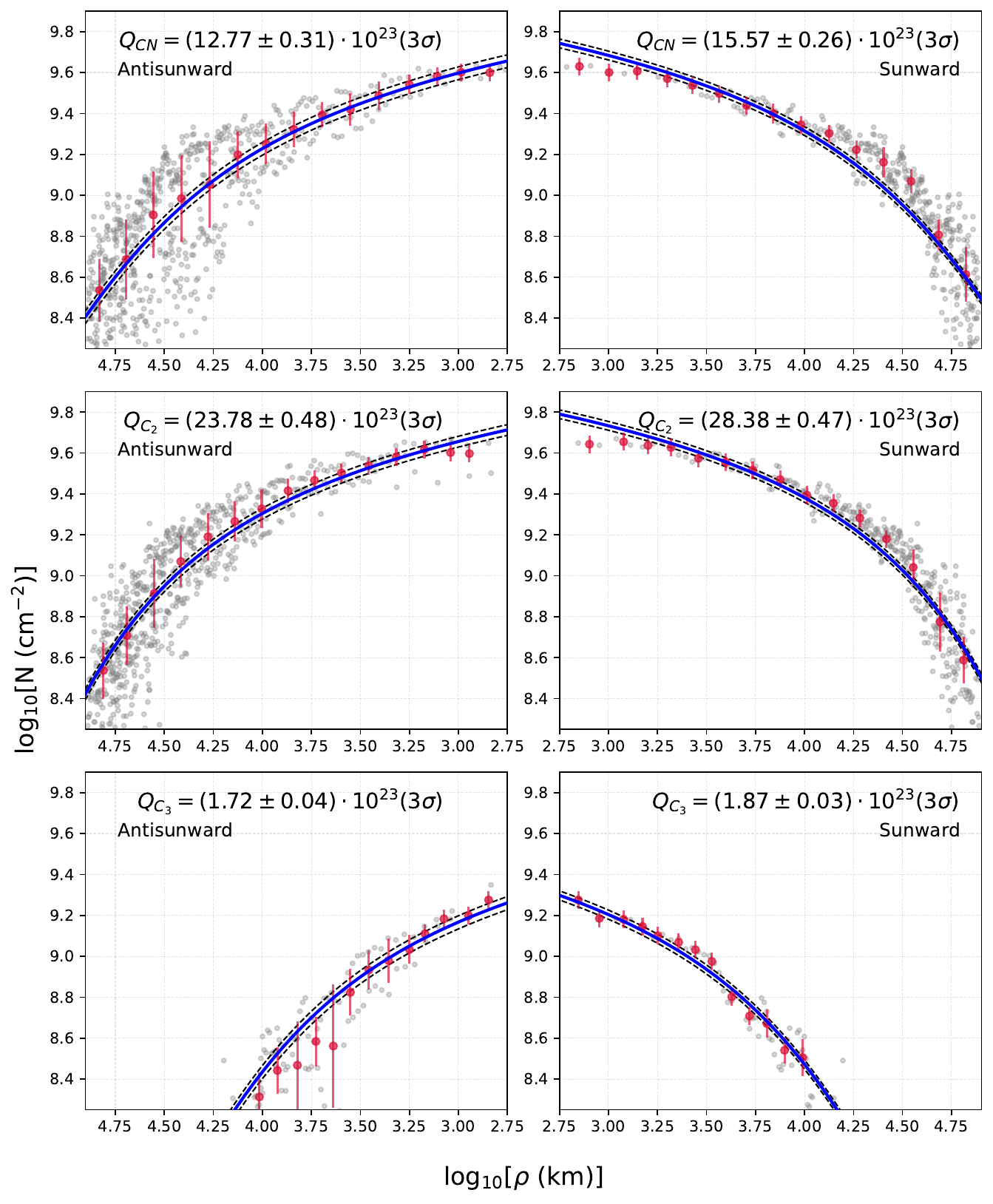}
    \caption{Column density profiles versus the projected cometocentric distance, both in log scale %Column density profiles in logarithmic scale, $\log_{10} N(\rho)$, versus the projected cometocentric distance $\rho$ also in log scale,
    for CN, C$_2$ and C$_3$ (from top to bottom) on May 13, 2021 at 03:11:12 . Grey dots represent the observational data and red circles with error bars show the data binned as described in the text. The best-fit Haser profiles for each species and direction are represented by thick blue lines, while dashed black lines indicate the associated uncertainties, based on the results presented in Table~\ref{tab:production_rates}.}
    \label{fig:7P_haser_profiles}   
\end{figure}

Table~\ref{tab:production_rates} shows best-fit production rates obtained for each night as derived from the fitting procedure of our stacked data for all species analysed in this work. Fig.~\ref{fig:7P_haser_profiles} shows, as an example, the results of the best Haser fit to the observed $\log N(\rho)$  for CN, C$_2$, and C$_3$, obtained from the stacked spectra of May 13, 2021, at 03:11:12~UT, which is the mid-time of the acquisition time of the sequence, with the slit oriented along the Sun-antisun direction. %The values derived for the sunward and antisunward directions of each %spectrum are also presented in Table~\ref{tab:production_rates}.\par
%However, this method of estimating uncertainty produced profiles that did not encompass most of the data contained within each bin. Therefore, we adjusted the uncertainty estimation according to this criterion: for each bin, we obtained two values corresponding to the previously calculated bin value, adjusted by adding and subtracting its associated uncertainty. This approach allowed us to generate "upper" and "lower" synthetic observational profiles. We then fit both profiles for each $Q$ value used in our simulations and selected the \( Q \) values corresponding to the lowest $\chi ^2$ values. We apply this procedure for the sunward and the antisunward directions, and for the combined data from both directions resulting in a single radial profile (labelled $Avg$) to obtain the average values in Table~\ref{tab:production_rates}.\par
%The computed gas production rate for every species analysed in this work can be seen in Table~\ref{tab:production_rates}. 
Averaged values for our dates would be 1.15, 2.32 and 0.17 $\times 10^{24}$ s$^{-1}$ for CN, C$_2$, and C$_3$, respectively.\par 

Concerning the production rates given in Table~\ref{tab:production_rates}, we observe that reproducing the enhanced column density in the sunward direction requires a larger gas production rate than in the opposite direction. This difference is larger than our three-sigma uncertainty, which is \textcolor{black}{reinforcing} an asymmetric gas emission in those directions in the case of CN and $\rm C_2$, \textcolor{black}{However, this asymmetry cannot be confirmed in the case of C$_3$ due to the 10\% uncertainty in the column density profiles}. \par 

We compare the production rates derived in this work with those of previous apparitions. To our knowledge, there are no previous estimates for both C$_2$ and C$_3$ for this comet.  Regarding CN, \cite{COCHRAN2012144} reported an upper limit for CN production of $ 5.75 \times 10^{ 23}$s$^{-1}$ during the 1989 apparition when the comet was at $r_h = 1.73$ AU. For the 2021 apparition, the CN production rate, based on our observations from May 8, May 13, and May 15 is $Q_{\text{CN}} = 1.15\times 10^{24}$~s$^{-1}$. We scale our value to $r_h = 1.73$~AU following a power law of -2.7 \citep{AHEARN1995223} yielding a CN production rate of $4.79 \times 10^{23}$s$^{-1}$. Thus, although both estimates correspond to different orbits, it could be said that our result is consistent with the upper limit reported by \cite{COCHRAN2012144}. Nevertheless, our estimate would be much lower than the measured one by \cite{1996ApJ...459..729F} at 1.42~AU during the 1989 apparition, about a factor of 5 scaling our value to that heliocentric distance using the aforementioned power law.  With regard to observations during the 2021 perihelion passage, our result also contrasts with the findings of \cite{Blain2022Dust} who reported a {\it factor of 2-4 more CN during the 2021 apparition than the upper limits} from \cite{COCHRAN2012144}, as said above. Unfortunately, there are no details of the observations in \cite{Blain2022Dust} and it is not possible to study the reasons after the discrepancies. Speculating, and given the frequent outbursts observed in this comet \citep{2022PSJ.....3..173L}, the discrepancies could be due to a potential variable activity in short time-scales.\par
%, suggesting that $Q_{\text{CN}}$ has remained stable despite the three perihelion passages between measurements. Such stability is reasonable given the small change in perihelion distance, from 1.26~AU in 1989 to 1.23~AU in 2021.\par %\textcolor{black}{Similarly, \cite{Blain2022Dust} detected a factor of $\sim$~2-4 more CN during 7P’s 2021 apparition than the 1989 upper limits measured by \cite{COCHRAN2012144}, which is also consistent with our results.}
% si lo calculo con las mismas scale lengths que usa cochran queda 5.57e23, que también es compatible con el resultado que ella obtenía. %'Dust Activity & Gas Emission in 7P/Pons-Winnecke': valor 2-4 veces mayor para Q(CN) que el upper limit the cochran, pero no especifican ni rh ni nada, entonces no se puede comparar. 
CN production rate of 7P seems to be very low when compared to other Jupiter family comets. Considering comets reported by \citet[\textcolor{black}{Fig.~11}]{COCHRAN2012144}, at the heliocentric distance of 1.25~AU, 7P would have a CN activity similar to \textcolor{black}{2P/Encke, and 67P/Churyumov-Gerasimenko} with the rest having much larger CN production rates. 
\textcolor{black}{It is necessary to note that the production rate of 67P for CN from \cite{COCHRAN2012144} has been estimated as $10^{24}~$s$^{-1}$ from the extrapolation of the various observations included in their study. We compare with this extrapolated value because we interpret it as the "average" behaviour of 67P. Somehow expected, this extrapolated "averaged" value is certainly lower than the more recent estimate by \cite{2017MNRAS.469S.222O} just near 67P perihelion ($6.72 \times 10^{24}$s$^{-1}$ - $10^{25}$s$^{-1}$), but still compatible within the dispersion shown in \cite{COCHRAN2012144}}

%\textcolor{black}{On the other hand, \citep{2017MNRAS.469S.222O} reported values ranging from $6.72 \times 10^{24}$s$^{-1}$ to 1 $10^{25}$s$^{-1}$ for the CN emission of 67P at 1.25~AU. We suggest that the origin of this discrepancy is because the results obtained by \cite{COCHRAN2012144} can be interpreted as representing an 'average' behavior of 67P, as they made no difference between hemispheres. In their work,  \cite{COCHRAN2012144} present fits of the activity of several comets as a function of heliocentric distance, which provides insight into their evolution. The fits are derived without distinguishing between the illuminated hemisphere, yielding a CN production rate for 67P of around $10^{24}$s$^{-1}$ at approximately 1.25 AU, values that align with our results.} %\textcolor{black}{Furthermore, when comparing the production rate of 7P from \citet[Table 4]{1996ApJ...459..729F}, we observe that 7P exhibits significantly lower emission than the other comets in the survey, even with a production rate that is about five times larger than ours.}

\textcolor{black}{Table~\ref{tab:production_rates} also provides the $\log(Q_{\text{C}_3}/Q_{\text{CN}})$ and $\log(Q_{\text{C}_2}/Q_{\text{CN}})$ratios, hereafter, C$_3$/CN and C$_2$/CN respectively.} We obtain values from C$_3$/CN ranging from -0.92 to -0.75, and C$_2$/CN varies between 0.26, observed on May 13 (sunward direction), and 0.38 on May 8 (both sunward and antisunward directions). Following the criteria used in \citet[\textcolor{black}{Table~6}]{AHEARN1995223} (C$_2$/CN $\geq -0.18$ for a 'typical' comet) we classify 7P as having typical abundances in terms of long-chain hydrocarbons. It is important to note that the taxonomic classification is highly sensitive to the scale lengths applied in the calculation of production rates. However, we employed consistent parameters to fit the Haser model to our observations.  %no solo > -0.18, para los parámetros que usan, 'typical' taxonomi if log(Q(C2)/Q(CN) > 0.06 (en el texto debajo de la tabla 6). Se cumple también. %A hearn et al: we restrict our taxonomic analysis to the gaseous species, because A f rho is a much more model-dependent measure of dust production than Q is for the gas species. 
With regard to the taxonomy by \cite{COCHRAN2012144}, we have recalculated C$_3$/CN and C$_2$/CN by using the scale lengths given in that reference. We then obtain C$_3$/CN  between -1.05 and -0.89, and  C$_2$/CN  between 0.19 and 0.33. \cite{COCHRAN2012144} define carbon depleted comets as those simultaneously satisfying  C$_3$/CN $\leq -0.86$ and  C$_2$/CN $\leq 0.02$. Thus, although 7P could be considered C$_3$ depleted because our measurements fulfill the condition, C$_2$/CN is much larger than 0.02. Therefore, and strictly, 7P cannot be considered as carbon-depleted. Both criteria, that of \cite{AHEARN1995223} and that of \cite{COCHRAN2012144}, would suggest that 7P  has a typical composition, although it can be considered C$_3$ depleted with regard to Jupiter family comets.\par

CN measurement and the relationship $\log(Q_{\text{OH}}/Q_{\text{CN}}) = 2.5$ found by \citep{AHEARN1995223} allow us to roughly estimate the water production rate of the comet near perihelion. %, we use the mean ratio of OH to CN,  $\log(Q_{\text{OH}}/Q_{\text{CN}}) = 2.5$  \citep{AHEARN1995223}, as no spectroscopic information on OH emission is available in our dataset.
%Using that relation, and the branching ratio of water photodissociation of 0.91 into OH, we calculate the water production rate. 
Assuming a branching ratio for water photodissociation of 0.91, the previous expression yields a water production rate for 7P of \textcolor{black}{3, 5 and 4~$\times 10^{26}$ s$^{-1}$ on May 8, 13 and 15 respectively, with an average water production rate of $4\times 10^{26}$ s$^{-1}$} over the three nights. The procedure is similar to that of~\cite{2001Icar..150..124L}, and our uncertainty would also be affected by a factor of about two.\par 
Water production rate can also be estimated by using the relationship $\log(Q_{\text{H}_2\text{O}}) = 30.675 - 0.2453 m_h$ obtained by \cite{2008LPICo1405.8046J}, which relates the magnitude of the coma, $m_h$\footnote{$ m_h$ denotes heliocentric magnitude, $m_h = m_v - 5\log \Delta$, with $m_v$ the visual magnitude of the coma and $\Delta$ the geocentric distance.}, to the water production rate. From our observations, the visual magnitude of the coma is estimated as \textcolor{black}{15.18, 15.03, and 14.85} for May 8, 13, and 15 respectively. Thus, the previous relationship gives a water production rate for 7P of \textcolor{black}{$(4.19 \pm 0.24) \times 10^{26}$ s$^{-1}$ for May~8, $(4.30 \pm 0.29) \times 10^{26}$ s$^{-1}$ for May~13, and $(4.66 \pm 0.30) \times 10^{26}$ s$^{-1}$ for May~15, with an average value of $(4.38 \pm 0.28) \times 10^{26}$ s$^{-1}$}, consistent with the value previously calculated from the CN production rate. %Furthermore, considering the branching ratio of water photodissociation into OH, the OH production rate turns into $4.91 \times 10^{26}$ s$^{-1}$, giving rise to $\log(Q_{\text{OH}}/Q_{\text{CN}})= 2.64 $. 
Indeed, if the water production derived from the visual magnitude is used to calculate the  $\log(Q_{\text{OH}}/Q_{\text{CN}})$ ratio, values between \textcolor{black}{2.44 and 2.65} are obtained, which agree well with the customary 2.5 in \citet{AHEARN1995223}. 
\textcolor{black}{By using the estimated water production rates from the visual magnitude, we obtain a water mass loss of 12.5 kg s$^{-1}$, 12.9 kg s$^{-1}$, and 13.9 kg s$^{-1}$ for the three relevant dates, respectively.}
\par 

All previous production rates, including that of water, indicate that 7P can be classified as a low-activity comet compared with other comets, as e.g. shown in~\cite{COCHRAN2012144}. This classification aligns with \cite{2021DPS....5321009B}, who described 7P as a "dry" comet. %The term dry typically refers to comets with a high dust-to-ice ratio in the nucleus, which could lead to low activity levels. However, since we measure the dust-to-gas ratio from the coma, it is important to note that this quantity does not necessarily reflect the dust-to-ice ratio in the nucleus, as pointed out by \cite{2020SSRv..216...44C}. 
We understand that "dry" refers to the nucleus that develops a dust mantle quenching sublimation processes. 
\textcolor{black}{7P is a short-period comet that has undergone approximately 30 apparitions since its discovery. After multiple perihelion passages, sublimation fronts might have receded from the nucleus surface, where an insulating dust layer forms. This could explain the low activity levels currently observed in 7P. This interpretation aligns with studies suggesting that only a small fraction of the surface—around 1.4\%—needs to be active to account for the observed activity levels \citep{2024SoSyR..58..456N}.
}

\section{Estimating the dust-to-gas ratio}\label{sect:dust-to-gas}
\textcolor{black}{Our Monte Carlo model provides a reasonably good estimate of the dust mass released from the nucleus as a function of the heliocentric distance (See Fig.~\ref{fig:masslossrate}). On May 8, 13 and 15, 7P produces approximately 20~kg~s$^{-1}$, 20.4~kg~s$^{-1}$ and 21.2~kg~s$^{-1}$ of dust, respectively. On the other hand, $A(\theta) f\rho$ \citep{1984AJ.....89..579A} parameter is commonly used to estimate the dust production rate~\citep[see e.g.][]{2024PSJ.....5...25G}. From our observations, we obtain $A(\theta) f\rho$  values of 22.93, 26.88, and 34.31 for May 8, 13, and 15, 2021.}\par 
\textcolor{black}{
Assuming the water production rate represents the gas production rate, the dust-to-gas mass ratios estimated with our Monte Carlo model in the coma of 7P are 1.6 for May 8 and 13, and 1.5 for May 15. On average, we find a value for dust-to-gas mass ratio of roughly 2. We compare the dust-to-gas mass ratio derived from $A(\theta) f\rho$  using the approach described by~\cite{2001Icar..150..124L} in line with \cite{AHEARN1995223}. The so obtained M$_\text{dust}$/M$_\text{gas}$ are 1.5, 1.7, and 2.0 for the three nights. Both methods consistently yield a dust-to-gas ratio of approximately 2 for 7P near perihelion.}\par 

To place this value into context, we compare it with the two other aforementioned JFCs—67P and 2P—that display similar CN production rates at comparable heliocentric distances. On the one hand, 67P, thanks to the Rosetta mission, is one of the best studied JFCs. However, as noted by \cite{2020SSRv..216...44C}, its dust-to-gas ratio remains elusive, showing a large dispersion depending on the method and data used to estimate it. The dust-to-gas mass ratio for 67P can be estimated as done in this work from the study performed by \cite{2017MNRAS.469S.186M}, who determined the dust mass loss from Monte Carlo modelling. From the retrieved dust mass loss and 67P water production \citep[as shown in][their Fig. 7]{2017MNRAS.469S.186M}, a dust-to-gas ratio of $\sim$~4.5 is obtained (a value within the range reported by \cite{2020SSRv..216...44C}). On the other, and unfortunately, it is not possible to estimate the dust-to-gas mass ratio for comet 2P/Encke as in the current study. However, literature provides us with consistent measurements of this ratio at $\sim$1.25 AU, see e.g. \cite{2001MNRAS.326..852S} and references therein, estimated to be smaller than 0.10, i.e. much smaller than that of 7P.

It is always tempting to try to connect the dust-to-gas ratio measured in the coma with nucleus characteristics. Nevertheless, the previous context indicates that it is difficult. Both comets, 67P and 2P, have very different dust-to-gas ratios. The former could be described as "dust rich" \citep[see e.g.][]{2017MNRAS.469S.475R} while the latter would be described as "gas rich" (or "dust poor"). 7P would be between the two of them. It seems that water production is not responsible for the differences in dust-to-gas ratio. Both comets, 67P and 2P, \citep[see e.g.][]{2001Icar..152..268M, 2020MNRAS.498.3995L} would have water production rates at least one order of magnitude larger than that of 7P (as estimated above) at 1.25~AU. %It has been suggested \citep{2017MNRAS.469S.276O} that most of the dust mass loss from 67P was in the form of comparatively large chunks, and recent thermophysical modelling by \cite{2020MNRAS.493.3690G} suggests that those large chunks can only be lifted by the effect of CO$_2$, not by the water. If those results are confirmed and can be generalized, a possible explanation, admittedly speculative, for the dust-to-gas ratio differences in the three comets mentioned above is that it could reflect the degree of CO$_2$ evolution. Nevertheless, this argument does not seem to be plausible as comet 2P, the one with lower dust-to-gas ratio, is cataloged as CO$_2$-rich \citep{2013Icar..226..777R}.

\section{Summary and conclusions} \label{sect:conclusions}
We present a comprehensive analysis of the dust and gas emission of comet 7P, with results derived from both image fitting and photometric modelling.  

The dust analysis was performed via our Monte Carlo simulator, using the images of the comet taken by the authors (pre-perihelion) and complemented with images from ZTF (pre- and post-perihelion). The best-fit dust ejection model provides key insights into the physical parameters governing the dust production. Notably, the peak production rate was determined to be 83 kg s\(^{-1}\), occurring approximately 15 days post-perihelion. The derived parameters are consistent with ‘typical’ cometary dust characteristics, such as a power-law size distribution index of \(-3.7\), and suggest a peak ejection velocity ranging from 3 m s\(^{-1}\) for largest particles to 23 m s\(^{-1}\) for the smallest ones.  

The isophotes of the coma are well shaped by the model, which reproduces the images observed over a broad interval (from 2021 Feb 15 to 2021 Jun 14, including both pre- and post-perihelion observations). However, deviations were observed during periods of unexpected activity, such as the outburst around 2021 June 2, previously reported by \cite{2024SoSyR..58..456N}. 

In addition to the dust, we analyzed the gas activity of 7P during its 2021 pre-perihelion passage at a heliocentric distance of approximately \(r_h \sim1.25\)~AU. Our spectra covered the pre-perihelion phase, with data collected on May 8, 13, and 15, and the perihelion was on May \textcolor{black}{27}. While the observations did not identify outbursts during this period, such activity is known to occur post-perihelion, as seen in the dust ejection models.\par
We derived the production rate for CN, C\(_2\) and  C\(_3\) using the Haser model with standard scale length parameters for parent and daughter molecules. All profiles, notably for CN and C\(_2\), showed significant asymmetries in the Sun-antisun directions, with higher production rates in the sunward direction, what reflects the heterogeneous nature of the cometary activity and the influence of solar-driven processes on volatile release.

The mean CN production rate obtained from our observations in 2021 is \(1.15 \times 10^{24}\) s\(^{-1}\). When scaled to the appropriate heliocentric distances, our estimate is consistent with the upper limit reported by \cite{COCHRAN2012144} in the 1989 apparition, but lower than the values reported by \cite{1996ApJ...459..729F} also for the 1989 apparition. Given the frequent number of outbursts that this comet suffers, this latter discrepancy could be attributed to potential variability in short time scales. %a localized outburst during the \cite{1996ApJ...459..729F}  observations, potentially caused by the exposure of a volatile-rich region on the surface, which could temporarily enhance CN emission through surface disruptions \textcolor{black}{ref(?)}, leading to a transient increase in CN production. 
Regarding other molecules, we also derived the C$_2$ and C$_3$ average production rates as 2.32$\times 10^{24}$ s$^{-1}$ and 1.69$\times 10^{23}$ s$^{-1}$, respectively. To our knowledge, there are no previous estimates of the production rates of these molecules. 

As for the ratios, our estimation of \textcolor{black}{C$_2$/CN}
%$\log(Q_{\text{C}_2}/Q_{\text{CN}})$ 
ranged from 0.26 to 0.38, and %$\log(Q_{\text{C}_3}/Q_{\text{CN}})$ 
\textcolor{black}{C$_3$/CN} ranged from \(-0.92\) to \(-0.75\). With these results, comet 7P is classified as having a typical composition based on the criteria of \citet{AHEARN1995223}. When the scale lengths of \cite{COCHRAN2012144} are considered, %$\log(Q_{\text{C}_2}/Q_{\text{CN}})$ 
\textcolor{black}{C$_2$/CN} ratio ranges from 0.19 to 0.33 while %$\log(Q_{\text{C}_3}/Q_{\text{CN}})$ 
\textcolor{black}{C$_3$/CN} varies from -1.05 to -0.89. Strictly, these values do not allow to classify 7P as carbon-depleted according to the two necessary conditions from \cite{COCHRAN2012144} but 7P seems to be somewhat C$_3$ depleted.

Regarding the dust-to-gas mass ratio, from the derived water production rates of $\sim$ 10 kg~s\(^{-1}\) and the dust loss values obtained from our Monte Carlo simulation $\sim$ 20 kg~s\(^{-1}\) it has been obtained that at 1.25~AU it is \textcolor{black}{of the order of 2.}%between 1.4 and 2.2.

\section*{Acknowledgements}

Based on observations obtained with the instrument CAFOS at the 2.2m telescope of the Calar Alto Observatory (CAHA), and with the Samuel Oschin Telescope 48-inch Telescope at the Palomar Observatory as part of the Zwicky Transient Facility project. ZTF is supported by the National Science Foundation under Grant No. AST-2034437 and a collaboration that includes Caltech, IPAC, Weizmann Institute for Science, Oskar Klein Center at Stockholm University, University of Maryland, Deutsches Elektronen-Synchrotron and Humboldt University, TANGO Consortium of Taiwan, University of Wisconsin at Milwaukee, Trinity College Dublin, Lawrence Livermore National Laboratories and IN2P3, France. Operations are carried out by COO, IPAC and UW.

We thank the amateur association \texttt{Cometas\_Obs} for providing us with photometric data for comet 7P.
The authors acknowledge financial support from grants PID2021-123370OB-I00, PID2021-126365NB-C21, and from the Severo Ochoa grant CEX2021-001131-S MICIU/AEI/ 10.13039/501100011033. I. Mariblanca-Escalona acknowledges financial support from the FPI grant PRE2022-105422 funded by MICIU/AEI/ 10.13039/501100011033 and by ESF+.
This research has made use of data provided by the International Astronomical Union's Minor Planet Center.
The Monte Carlo dust tail code makes use of the JPL Horizons ephemeris generator system.

\section*{DATA AVAILABILITY}
The data underlying this article will be shared on reasonable request to the corresponding author.

%%%%%%%%%%%%%%%%%%%% REFERENCES %%%%%%%%%%%%%%%%%%

% The best way to enter references is to use BibTeX:

\bibliographystyle{mnras}
\bibliography{references} % if your bibtex file is called example.bib

% Alternatively you could enter them by hand, like this:
% This method is tedious and prone to error if you have lots of references
%\begin{thebibliography}{99}
%\bibitem[\protect\citeauthoryear{Author}{2012}]{Author2012}
%Author A.~N., 2013, Journal of Improbable Astronomy, 1, 1
%\bibitem[\protect\citeauthoryear{Others}{2013}]{Others2013}
%Others S., 2012, Journal of Interesting Stuff, 17, 198
%\end{thebibliography}

%%%%%%%%%%%%%%%%%%%%%%%%%%%%%%%%%%%%%%%%%%%%%%%%%%

%%%%%%%%%%%%%%%%% APPENDICES %%%%%%%%%%%%%%%%%%%%%

%\appendix

%\section{Some extra material}

%If you want to present additional material which would interrupt the flow of the main paper, it can be placed in an Appendix which appears after the list of references.

%%%%%%%%%%%%%%%%%%%%%%%%%%%%%%%%%%%%%%%%%%%%%%%%%%

% Don't change these lines
\bsp	% typesetting comment
\label{lastpage}
\end{document}